\documentclass[rnote]{aa}
\usepackage{graphicx}
\usepackage{color}
\usepackage{amssymb}
\usepackage{amsmath}
\usepackage{xspace}
\usepackage{xargs}
\usepackage[varg]{txfonts}

\newcommand{\B}[1]{{{#1}}}

\newcommand{\myemail}{Rebecca.Kennedy@nottingham.ac.uk}

\newcommand{\galfitm}{{\scshape galfitm}\xspace}
\newcommand{\megamorph}{{MegaMorph}\xspace}
\newcommand{\sersic}{S\'ersic\xspace}

\newcommandx{\N}[2][1= ,2= ]{$\mathcal{N}^{#1}_{#2}$\xspace} 
\newcommandx{\R}[2][1= ,2= ]{$\mathcal{R}^{#1}_{#2}$\xspace}
\newcommand{\re}{$R_{\rm e}$\xspace}
\newcommandx{\Rss}{$\mathcal{R}_{\rm ss}$\xspace}
\newcommandx{\Nss}{$\mathcal{N}_{\rm ss}$\xspace}

\begin{document}
\titlerunning{Galaxy colour gradients}
\title{Galaxy And Mass Assembly (GAMA): Galaxy colour gradients versus colour, structure and luminosity}
\authorrunning{Kennedy et al.}
\author{Rebecca~Kennedy\inst{\ref{inst1}}\thanks{E-mail: \myemail}\and Steven~P.~Bamford\inst{\ref{inst1}}\and Boris~H\"au\ss ler\inst{\ref{inst2},\ref{inst3},\ref{inst4}}\and Sarah~Brough\inst{\ref{inst5}}\and Benne~Holwerda\inst{\ref{inst6}}\and Andrew~M.~Hopkins\inst{\ref{inst6}}\and Marina~Vika\inst{\ref{inst7}}\and Benedetta~Vulcani\inst{\ref{inst8}}}

\institute{School of Physics \& Astronomy, The University of Nottingham, University Park, Nottingham, NG7 2RD, UK\label{inst1}\and University of Oxford, Denys Wilkinson Building, Keble Road, Oxford, Oxon OX1 3RH, UK\label{inst2}\and University of Hertfordshire, Hatfield, Hertfordshire AL10 9AB, UK\label{inst3}\and European Southern Observatory, Alonso de Cordova 3107, Vitacura, Santiago, Chile\label{inst4}\and Australian Astronomical Observatory, PO Box 915, North Ryde, NSW 1670, Australia \label{inst5}\and University of Leiden, Sterrenwacht Leiden, Niels Bohrweg 2, NL-2333 CA Leiden, The Netherlands\label{inst6}\and IAASARS, National Observatory of Athens, GR-15236 Penteli, Greece\label{inst7}\and School of Physics, University of Melbourne, VIC 3010, Australia\label{inst8}}

\abstract{Using single-component fits to SDSS/UKIDSS images of galaxies in the G09 region of the GAMA survey we study radial colour gradients across the galaxy population. We use the multiwavelength information provided by \megamorph analysis of galaxy light profiles to calculate intrinsic colour gradients, and divide into six subsamples split by overall \sersic index ($n$) and galaxy colour.

We find a bimodality in the colour gradients of high- and low-$n$ galaxies in all wavebands, which varies with overall galaxy luminosity. Global trends in colour gradients therefore result from combining the contrasting behaviour of a number of different galaxy populations.
The ubiquity of strong negative colour gradients supports the picture of inside-out growth through gas accretion for blue, low-$n$ galaxies, and through dry minor mergers for red, high-$n$ galaxies. An exception is the blue high-n population, with properties indicative of dissipative major mergers.
}


\maketitle

\section{Introduction}

The global colours of galaxies can be driven by a number of factors, including their dust content, the average age of their stellar populations and metallicity. The colour within a given galaxy also often varies radially, reflecting \B{internal trends in the age, metallicity and dust extinction of its stellar populations}. 
The existence of negative colour gradients (i.e. redder in the centre and bluer in the outskirts) in elliptical galaxies is well-documented \citep[see][and references therein]{Peletier1990,Gonzalez-Perez2011}. \citet{LaBarbera2010} find that these colour gradients can be linked to the overall colour and luminosity of an early-type galaxy (ETG); steeper negative colour gradients are more commonly found in bluer or more luminous ETGs. Steep negative colour gradients have also been observed in late-type galaxies (LTGs) and are thought to indicate the presence of young stars in their outer regions \citep{Gonzalez-Perez2011}. There are, however, inherent difficulties in measuring consistently accurate stellar population colours in LTGs due to interstellar extinction caused by dust. This dust extinction is most problematic in optical wavebands \citep[see][and references therein]{deJong1994}. Moving to longer wavelengths combats this problem, as observations in the near infra red (NIR) are less affected by dust. The NIR is also better able to observe the older stellar populations within a galaxy, which contain most of the stellar mass \citep{deJong1994}.

Recently, advances have been made in using colour information to constrain age and metallicity gradients in ETGs \citep[e.g.][]{LaBarbera2010,Carter2011}.
Fewer studies of a similar nature have been attempted for spiral galaxies, partly due to their increased dust content compared to elliptical galaxies, and partly due to their more complex structure.
\cite{Bell2000} find that the inner regions of most low-inclination spiral galaxies are older and more metal-rich than their outskirts which supports inside-out formation, although age gradients do not necessarily correlate with metallicity gradients \citep[e.g.][and references therein]{Tortora2010,Sanchez2014}. Similarly, $(g-r)$ and $(g-i)$ rest-frame global colours are mainly related to the age of the galaxy, and do not depend strongly on metallicity \citep{Gonzalez-Perez2011}.


In a previous work \citep{Kennedy2015} we studied the wavelength dependence of galaxy structure to reveal the connections between the stellar populations within a galaxy and their spatial distributions.  We found evidence for multi-component structure in many galaxies, even those visually classified as elliptical (e.g., \citealt{Kennedy2016}). This suggests that considering structural variations with wavelength may provide fundamental insights into galaxy formation. Nevertheless, studying internal colour gradients is a complementary and widely-used approach, and thus it is important to understand how our results on the wavelength-dependence of structure relate to more traditional colour gradients.
In this Research Note we study how colour gradient varies with wavelength (from  $\nabla_{g-r}$ to  $\nabla_{g-H}$) for six subsamples split by overall \sersic index and galaxy colour, and then consider the luminosity dependence of these colour gradients.

\section{Data}
\label{sec:data}
Our sample comes from the G09 region of GAMA II \citep{Driver2009,Driver2011,Liske2015}, and is volume-limited to $M_{r} < -19.48$ mag and $z < 0.15$ when studying trends with luminosity (which gives a larger luminosity range but smaller sample of 5317 galaxies, with a stellar mass limit of $\sim 10^{9}M_{\odot}$), and $M_{r} < -21.2$ mag and $z < 0.3$ elsewhere (which gives a larger sample of 13825 galaxies at the expense of a smaller luminosity range with a stellar mass limit of $\sim 10^{10}M_{\odot}$). \galfitm \citep{Haussler2013,Vika2013} fits a single-\sersic wavelength-dependent model to all bands simultaneously, and returns (amongst other parameters) magnitude ($m$), \sersic index ($n$), and effective radius (\re). For more information on the data used here, see \citet{Haussler2013}; \citet[][hereafter V14]{Vulcani2014}; and \citet{Kennedy2015}.

As in V14 and \citet{Kennedy2015} we divide our sample into high- and low-\sersic index at $n_{r} = 2.5$, and separate $red$ and blue galaxies using a $u-r = 2.1$ colour cut. We then separate the blue galaxies into $green$ and $blue$ at $u-r = 1.6$ in an attempt to separate out the bluest, potentially starburst, population. We note that our $green$ population corresponds to the main population of star-forming galaxies, not specifically the green valley.

In our previous work we measured the wavelength dependence of \sersic index, $\mathcal{N} = n(H)/n(g)$, and effective radius, $\mathcal{R} = R_{\rm e}(H)/R_{\rm e}(g)$ between the $H$- and $g$-bands. We saw that there is a striking difference in the behavior of high-$n$ ($n_{r} > 2.5$) and low-$n$ ($n_{r} < 2.5$) galaxies; high-$n$ galaxies show a large change in \re (\R $<1$) but little change in n (\N $\sim 1$), whereas in low-$n$ galaxies show little change in \re with wavelength but a large change in $n$ with wavelength. Although \N and \R are useful for considering the dependence of structure on wavelength, they are hard to compare to the literature. Here we therefore convert these measurements into more traditional colour gradients which describe a change in stellar population with radius, rather than a change in radial profile with wavelength.

We derive our colour gradients using the logarithmic slope of a galaxy's radial colour profile between 0.1\re and 1\re \citep[see e.g. ][]{Peletier1990,LaBarbera2010}, taking \re as the effective radius in the $r$-band. The colours as a function of radius are obtained from our multi-band \sersic model fit to each galaxy. Both the \sersic index and effective radius of a galaxy are allowed to vary quadratically with wavelength during the fit. As we measure the colour gradient over a radial range defined in a single band (the $r$-band), both the variations of $n$ and \re with wavelength contribute to our measured colour gradients. We then use orthogonal regression to find $\nabla_{g-x}$, the gradient of the best-fitting profile where $x = rizYJH$.
Our large sample size allows us to study both early- and late-type galaxies in 7 wavebands. We can therefore provide a more complete picture of how colour gradients - and hence stellar mass growth channels - vary across the galaxy population. We are also able to extend the work of \citealt{Kennedy2015} to study in detail the luminosity dependence colour gradients for our subsamples of galaxies.

\section{Results}

\begin{figure}
	\centering
	\includegraphics[width=0.45\textwidth]{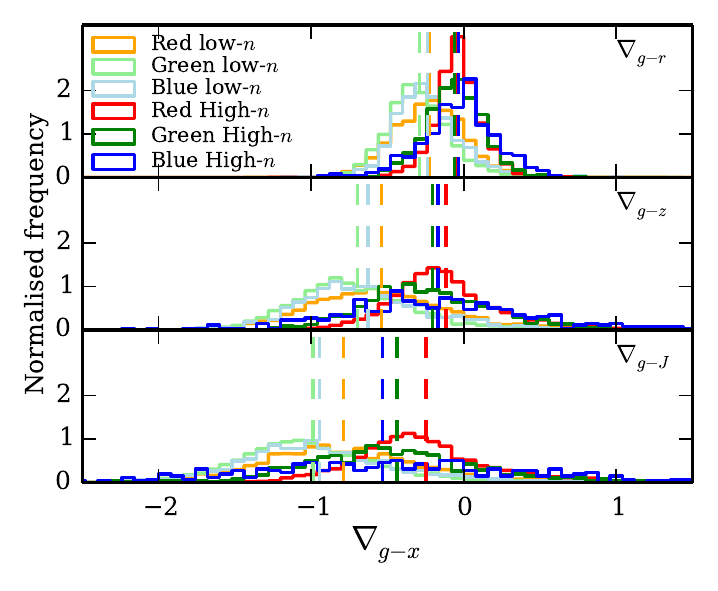}
	\caption{Distributions of $\nabla_{g-x}$ for galaxies in each of our colour/\sersic index subsamples (see section~\ref{sec:data} for more details), normalised to unit integral. Each panel shows the distribution for a different pair of bands. The median of each distribution is indicated by a vertical dashed line. These median values, and their uncertainties, can be found in Table \ref{table:Table1}. A bimodality in colour gradients for high- and low-$n$ galaxies can be seen in all wavebands.
	\label{fig:Fig1}}
\end{figure}

\begin{figure*}
	\centering
    \includegraphics[width=0.8\textwidth]{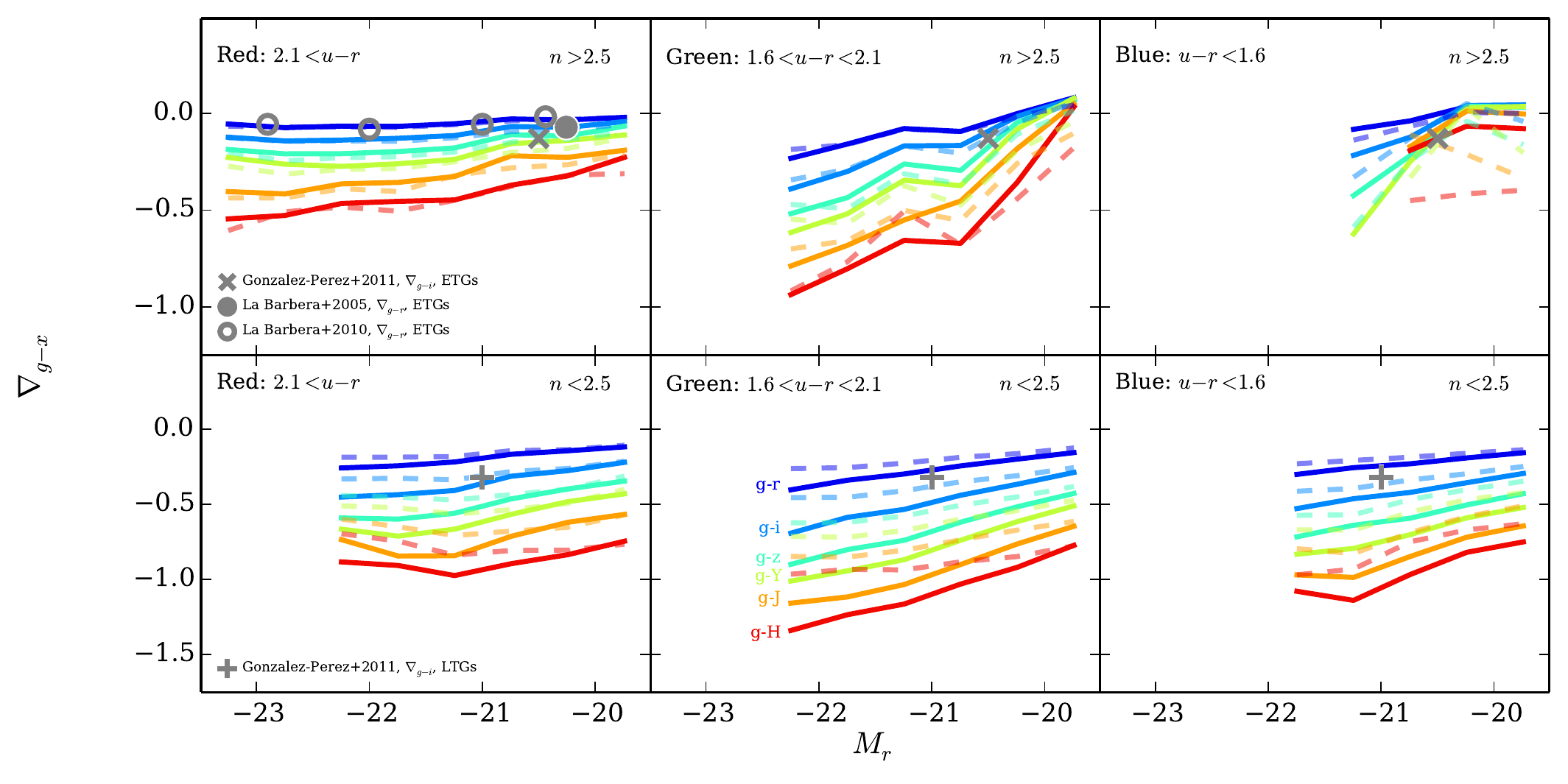}
	\caption{Median colour gradient, $\nabla_{g-x}$, where $x = rizYJH$, as a function of $r$-band absolute magnitude, $M_{r}$, in each of our colour/\sersic index subsamples. Solid and dashed lines show colour gradients over the radius ranges $0.1$\re--$1$\re, and $0.1$\re--$2$\re, respectively. Lines are limited to magnitude bins containing at least 10 objects. They are therefore only shown where each subsample makes up a significant fraction ($\gtrsim $1\%) of the galaxy population. Different colours correspond to different wavebands, with $g-r$ at the top to $g-H$ at the bottom, as indicated. Results of previous studies are overlaid in the panels which most closely match the subsample of that study (see legends).
    We see that only the red, high-$n$ population has galaxies brighter than $M_{r} \sim -22.5$, so these galaxies dominate at the bright end.
	\label{fig:Fig3}}
\end{figure*}

The variation in colour gradient with wavelength is summarised in Fig.~\ref{fig:Fig1}; we see that the majority of colour gradients are negative, i.e. appear bluer at larger radii. We have also examined the redshift dependence of these colour gradients in a similar manner to \cite{Kennedy2015}, and find that they do not change significantly. As we would expect, colour gradients are stronger for more widely spaced waveband pairs. We see a distinct difference in the colour gradients of low- and high-$n$ galaxies, regardless of overall galaxy colour. Our low-$n$ samples consistently have the strongest colour gradients. In \cite{Kennedy2016} we found these low-$n$ samples to contain galaxies with more significant disc components (i.e. a lower bulge/total flux ratio) than their high-$n$ counterparts, which is in agreement with \cite{Gonzalez-Perez2011} who find that steep negative gradients are more likely to be found in late-types. We can also see that the $red$ subsamples consistently have shallower colour gradients than the $blue$ and $green$ subsamples; once again we are in agreement with the finding of \cite{Gonzalez-Perez2011} that the redder a galaxy, the shallower its colour gradient. 
Our red, high-$n$ sample is a reasonable proxy for early-type galaxies, and the mean colour gradients we find (given in Table \ref{table:Table1} for all our subsamples) agree well with the colour gradients found in \citet{LaBarbera2010} for their sample of ETGs over the same range of wavebands.  These values are later shown in Fig.~\ref{fig:Fig3}. Within the uncertainties, the mean colour gradients of our low-$n$ samples also agree with those of \cite{Taylor2005} for `mid-type' (Sc) spiral galaxies, and \cite{Gonzalez-Perez2011} for late-type galaxies in a similar magnitude range.

\begin{table*}
\centering
\scalebox{0.8}{
	\begin{tabular}{ l | c  c  c  c  c  c  c  c  c  c  c  c  c  c}
		& \multicolumn{2}{c}{Red, low-$n$} & \multicolumn{2}{c}{Green, low-$n$} & \multicolumn{2}{c}{Blue, low-$n$} & \multicolumn{2}{c}{Red, high-$n$} & \multicolumn{2}{c}{Green, high-$n$} & \multicolumn{2}{c}{Blue, high-$n$} & max.\@ err. \\
        & $\mu$ & $\sigma$ & $\mu$ & $\sigma$ & $\mu$ & $\sigma$ & $\mu$ & $\sigma$ & $\mu$ & $\sigma$ & $\mu$ & $\sigma$ & on $\mu$\\ \hline
		$\nabla_{g-r}$   	& -0.22 & 0.24 & -0.29 & 0.21 & -0.24 & 0.23 & -0.04 & 0.16 & -0.06 & 0.19 & -0.03 & 0.25 & 0.01 \\
		$\nabla_{g-i}$		& -0.39 & 0.40 & -0.51 & 0.35 & -0.44 & 0.38 & -0.07 & 0.28 & -0.12 & 0.35 & -0.07 & 0.45 & 0.03 \\
        $\nabla_{g-z}$  	& -0.54 & 0.50 & -0.70 & 0.45 & -0.63 & 0.51 & -0.12 & 0.40 & -0.21 & 0.51 & -0.17 & 0.72 & 0.04 \\
        $\nabla_{g-Y}$   	& -0.64 & 0.54 & -0.82 & 0.50 & -0.76 & 0.57 & -0.16 & 0.46 & -0.28 & 0.60 & -0.29 & 0.92 & 0.05 \\
		$\nabla_{g-J}$  	& -0.79 & 0.58 & -0.99 & 0.54 & -0.95 & 0.62 & -0.25 & 0.52 & -0.44 & 0.71 & -0.53 & 1.20 & 0.07 \\
        $\nabla_{g-H}$  	& -0.95 & 0.61 & -1.12 & 0.55 & -1.11 & 0.68 & -0.40 & 0.52 & -0.67 & 0.80 & -0.79 & 1.44 & 0.08 \\
	\end{tabular}}
	\caption{Mean colour gradients, $\mu$, and standard deviations,  $\sigma$, for each of our colour/\sersic index subsamples. \B{The final column gives the maximum uncertainty on the mean values ($\sigma/\sqrt{N}$), corresponding to the smallest subsample, blue high-$n$.  For the larger subsamples, this uncertainty is up to an order of magnitude smaller.}
		\label{table:Table1}}
\end{table*}

As we saw in Fig.~\ref{fig:Fig1}, different galaxy types (in this context, different $n$ and colours) display contrasting colour gradients. We therefore show in Fig.~\ref{fig:Fig3} how colour gradients, $\nabla_{g-x}$ where $x = rizYJH$, change with luminosity depending on a galaxy's overall $u-r$ colour and \sersic index, over two radial ranges.

The variety of trends seen in each panel of Fig.~\ref{fig:Fig3} shows that the overall trends seen in the literature are actually the result of a number of different populations. Red, high-$n$ galaxies dominate at bright magnitudes (i.e. there are only red, high-$n$ galaxies brighter than $M_{r} \sim -22.5$) whilst low-$n$ galaxies dominate at faint magnitudes.
We note that colour gradients in high-$n$ galaxies do not change when measuring over a wider radius range (as previously found by \cite{LaBarbera2009} for ETGs, and shown as dashed lines in Fig.~\ref{fig:Fig3}). Low-$n$ galaxies do show slightly flatter colour gradients between $0.1$\re--$2$\re which suggests that bright, low-n galaxies have stronger colour gradients towards their centres. This is consistent with the presence of a bulge within a relatively homogeneous outer disc, with the bulge being more significant in brighter galaxies.

\citet{LaBarbera2010} find a `double-valued' behavior in the $\nabla_{g-r}$ colour gradient with luminosity for their early-type galaxies. We can see in Fig.~\ref{fig:Fig3} that this double-valued behavior can actually be considered a combination of the trends seen in the low- and high-$n$ $red$ populations. The colour gradients seen in a population are sensitive to the selection criteria used to define that population.

Our $red$, high-$n$ sample shows little change in colour gradient with luminosity in optical wavebands, but at longer wavelengths the faintest galaxies have the shallowest colour gradients. \cite{denBrok2011} study a sample of 142 ETGs and find that the colour gradients of elliptical galaxies (shown in \B{\citealt{Kennedy2016}} to be analagous to our red, high-$n$ population) become shallower at fainter magnitudes, which is consistent with our results. Note that only the red, high-$n$ population has a significant number of galaxies brighter than $M_{r} \sim -22.5$ mag, so these galaxies dominate the brightest end of the sample.
\citet{Tortora2010} use $\nabla_{g-i}$ as a measure of colour gradient, and find that colour gradients decrease with mass; i.e. more massive, and therefore more luminous, systems have the weakest colour gradients. This is consistent with our assumption that our high-$n$ populations contain elliptical galaxies, which tend to be larger and more massive than spirals and \B{S0s} \citep{Kelvin2012,Moffett2016}.

On the whole, the low-$n$ subsamples show little change in colour gradient vs. luminosity with wavelength (i.e. lines are approximately parallel to one another). However, in our high-$n$ subsamples we see that the faintest galaxies show very similar colour gradients, whilst the brightest objects show a larger change in colour gradient with wavelength. \cite{LaBarbera2010}, who study a sample of ETGs which are most likely to correspond to our high-$n$ subsamples, also find that the behavior of colour gradients with luminosity depends on both the wavebands in which the colour gradient was calculated, and the waveband in which the luminosity was determined.

Overplotted in grey in Fig.~\ref{fig:Fig3} we show the colour gradients for samples from some previous studies, in the panel to which their sample is most closely related. In the red high-$n$ panel we show $\nabla_{g-r}$ for \cite{LaBarbera2005}'s sample of luminous ETGs in clusters, which lies within the standard deviation (not shown here, but can be inferred from Table \ref{table:Table1}) of the faint end of our $\nabla_{g-r}$ line. In this panel we also show with open circles $\nabla_{g-r}$ for a larger sample of ETGs with $ugrizYJHK$ photometry from \cite{LaBarbera2010}. The $\nabla_{g-r}$ luminosity dependence agrees well with our red, high-$n$ sample, but sits a little lower at intermediate magnitudes. This is likely due to the ETG sample in \cite{LaBarbera2010} also containing galaxies which lie in our red, low-$n$ sample, which have stronger colour gradients than their high-$n$ counterparts at $M_{r} \sim -22$. The $\nabla_{g-i}$ colour gradient for \cite{Gonzalez-Perez2011}'s ETG sample lies directly on our $\nabla_{g-i}$ line at that $r$-band magnitude, whilst in the red, low-$n$ panel we plot \cite{Gonzalez-Perez2011}'s $\nabla_{g-i}$ for their brightest late-type sample which again agrees with our $\nabla_{g-i}$ for more discy objects.

\section{Conclusions}

We have measured \B{radial gradients in six colours} using multi-wavelength single-S{\' e}rsic model fits to galaxies in the GAMA survey.  These complement our previous studies \citep{Vulcani2014,Kennedy2015} on the variation of structural parameters with wavelength (\N and \R).  Our measurements correspond well to those in the literature, supporting the reliability of our methods and our earlier conclusions regarding the wavelength-dependence of galaxy structure.

We find that galaxies with differing S{\' e}rsic index and total colour display contrasting behaviour in terms of both the distributions and luminosity dependence of their colour gradients.  This means that care must be taken when comparing results for different sample selections.

The ubiquitous negative gradients we find are indicative of stellar populations becoming younger and/or lower metallicity as one progresses from the centres to the outskirts of galaxies.  While this is consistent with simple models of galaxy formation via early dissipative collapse (e.g. \citealt{Pipino2010}), in our $\Lambda$CDM Universe we expect the colour gradients of most galaxies to reflect their long-term hierarchical growth.

Strong negative gradients in low-$n$ systems are consistent with inside-out disc formation via gradual accretion of gas to their outskirts, via smooth flows or (gas-rich) minor mergers (e.g., \citealt{Lemonias2011,Wang2011}).
Such gradients are present in blue, low-luminosity galaxies, with little bulge component; they must therefore arise from stellar population and/or dust gradients within the discs themselves.  Nevertheless, the presence of a substantially redder bulge in more luminous disc galaxies \citep{Kennedy2016}, appears to further enhance colour gradients, particularly within 1\re.

The presence of \B{gradients, even if weak,} in bright, high-$n$ systems argues against these galaxies being the direct result of major mergers, since such violent interactions should flatten colour gradients \citep[see][and references therein]{Kim2013}.  Instead, their gradients may either result from the fading of bluer and/or lower-$n$ galaxies (e.g., \citealt{Bell2006}), or from a reintroduction of metal-poor and/or younger stellar populations in the outskirts of these galaxies, for example via dry minor mergers \citep{Eliche2013}.  The latter can also account for the rapid size evolution of passive galaxies \citep{Oser2010}.

\B{In contrast, lower-luminosity high-$n$ galaxies with bluer colours display very flat, or even slightly positive, colour gradients, especially in the optical.  We infer that this population is experiencing a period of central star-formation. These galaxies account for $2.4$\% of our sample, which corresponds well with the local major merger fraction: also $\sim 2$\% across a wide range of galaxy masses \citep[see][and references therein]{Khochfar2001,Keenan2014}. At the luminosities of interest, major mergers are mostly mixed elliptical-spiral interactions, and hence gas-rich \citep[see e.g.][]{Khochfar2003,Darg2010}. 
We therefore associate the blue high-$n$ population with dissipative major mergers.  However, less violent causes of central star formation, such as bar instabilities and minor mergers \citep[see e.g.][and references therein]{Kormendy2004}, may also contribute.
The rarity of these examples suggests their descendants will re-acquire negative colour gradients, via fading of the central star-formation and subsequent accretion. }

The colour profiles of most galaxies therefore result from continued accretion, with the ratio of gas to stars in the incoming material varying from high: in the case of lower-luminosity, lower-$n$ and bluer galaxies; to low: for higher-luminosity, higher-$n$ and redder galaxies.

Possible applications for our colour gradient measurements include multiwavelength strong lensing (information about the likely internal colour gradient of a galaxy will allow tighter constraints on mass models), and comparison with galaxy simulations (colour gradients provide more information with which to challenge models).  Our future work will use decomposed bulge and disc properties to further study the relationships between stellar populations and galaxy structure.  In particular, we will expand upon the present study by using multi-band method to measuring colour gradients for individual galaxy components, e.g. bulges and discs, separately.

\section{Acknowledgements}
R.K. acknowledges support from the Science and Technology Facilities Council (STFC).
B.V. acknowledges the support from an Australian Research Council Discovery Early Career Researcher Award (PD0028506).
\bibliographystyle{aa}
\bibliography{4Dec14}

\begin{thebibliography}{34}
\expandafter\ifx\csname natexlab\endcsname\relax\def\natexlab#1{#1}\fi

\bibitem[{{Bell} \& {de Jong}(2000)}]{Bell2000}
{Bell}, E.~F. \& {de Jong}, R.~S. 2000, MNRAS, 312, 497

\bibitem[{{Bell} {et~al.}(2006){Bell}, {Naab}, {McIntosh}, {Somerville},
  {Caldwell}, {Barden}, {Wolf}, {Rix}, {Beckwith}, {Borch}, {H{\"a}ussler},
  {Heymans}, {Jahnke}, {Jogee}, {Koposov}, {Meisenheimer}, {Peng}, {Sanchez},
  \& {Wisotzki}}]{Bell2006}
{Bell}, E.~F., {Naab}, T., {McIntosh}, D.~H., {et~al.} 2006, ApJ, 640, 241

\bibitem[{{Carter} {et~al.}(2011){Carter}, {Pass}, {Kennedy}, {Karick}, \&
  {Smith}}]{Carter2011}
{Carter}, D., {Pass}, S., {Kennedy}, J., {Karick}, A.~M., \& {Smith}, R.~J.
  2011, MNRAS, 414, 3410

\bibitem[{{Darg} {et~al.}(2010){Darg}, {Kaviraj}, {Lintott}, {Schawinski},
  {Sarzi}, {Bamford}, {Silk}, {Proctor}, {Andreescu}, {Murray}, {Nichol},
  {Raddick}, {Slosar}, {Szalay}, {Thomas}, \& {Vandenberg}}]{Darg2010}
{Darg}, D.~W., {Kaviraj}, S., {Lintott}, C.~J., {et~al.} 2010, MNRAS, 401, 1043

\bibitem[{de~Jong \& van~der Kruit(1994)}]{deJong1994}
de~Jong, R. \& van~der Kruit, P. 1994, Astronomy and Astrophysics Suppl. 106

\bibitem[{den Brok {et~al.}(2011)den Brok, Peletier, Valentijn, Balcells,
  Carter, Erwin, Ferguson, Goudfrooij, Graham, Hammer, Lucey, Trentham,
  Guzm\'{a}n, Hoyos, {Verdoes Kleijn}, Jogee, Karick, Marinova, Mouhcine, \&
  Weinzirl}]{denBrok2011}
den Brok, M., Peletier, R.~F., Valentijn, E.~A., {et~al.} 2011, MNRAS, 414,
  3052

\bibitem[{Driver {et~al.}(2011)Driver, Hill, Kelvin, Robotham, Liske, Norberg,
  Baldry, Bamford, Hopkins, Loveday, Peacock, Andrae, Bland-Hawthorn, Brough,
  Brown, Cameron, Ching, Colless, Conselice, Croom, Cross, {De Propris}, Dye,
  Drinkwater, Ellis, Graham, Grootes, Gunawardhana, Jones, van Kampen,
  Maraston, Nichol, Parkinson, Phillipps, Pimbblet, Popescu, Prescott,
  Roseboom, Sadler, Sansom, Sharp, Smith, Taylor, Thomas, Tuffs, Wijesinghe,
  Dunne, Frenk, Jarvis, Madore, Meyer, Seibert, Staveley-Smith, Sutherland, \&
  Warren}]{Driver2011}
Driver, S.~P., Hill, D.~T., Kelvin, L.~S., {et~al.} 2011, MNRAS, 413, 971

\bibitem[{Driver {et~al.}(2009)Driver, Norberg, Baldry, Bamford, Hopkins,
  Liske, Loveday, \& Peacock}]{Driver2009}
Driver, S.~P., Norberg, P., Baldry, I.~K., {et~al.} 2009, Astronomy \&
  Geophysics, 50, 5.12

\bibitem[{{Eliche-Moral} {et~al.}(2013){Eliche-Moral},
  {Gonz{\'a}lez-Garc{\'{\i}}a}, {Aguerri}, {Gallego}, {Zamorano}, {Balcells},
  \& {Prieto}}]{Eliche2013}
{Eliche-Moral}, M.~C., {Gonz{\'a}lez-Garc{\'{\i}}a}, A.~C., {Aguerri},
  J.~A.~L., {et~al.} 2013, A\&A, 552, A67

\bibitem[{Gonzalez-Perez {et~al.}(2011)Gonzalez-Perez, Castander, \&
  Kauffmann}]{Gonzalez-Perez2011}
Gonzalez-Perez, V., Castander, F.~J., \& Kauffmann, G. 2011, MNRAS, 411, 1151

\bibitem[{H\"{a}u{\ss}ler {et~al.}(2013)H\"{a}u{\ss}ler, Bamford, Vika, Rojas,
  Barden, Kelvin, Alpaslan, Robotham, Driver, Baldry, Brough, Hopkins, Liske,
  Nichol, Popescu, \& Tuffs}]{Haussler2013}
H\"{a}u{\ss}ler, B., Bamford, S.~P., Vika, M., {et~al.} 2013, MNRAS, 430, 330

\bibitem[{{Keenan} {et~al.}(2014){Keenan}, {Foucaud}, {De Propris}, {Hsieh},
  {Lin}, {Chou}, {Huang}, {Lin}, \& {Chang}}]{Keenan2014}
{Keenan}, R.~C., {Foucaud}, S., {De Propris}, R., {et~al.} 2014, ApJ, 795, 157

\bibitem[{Kelvin {et~al.}(2012)Kelvin, Driver, Robotham, Hill, Alpaslan,
  Baldry, Bamford, Bland-Hawthorn, Brough, Graham, H\"{a}ussler, Hopkins,
  Liske, Loveday, Norberg, Phillipps, Popescu, Prescott, Taylor, \&
  Tuffs}]{Kelvin2012}
Kelvin, L.~S., Driver, S.~P., Robotham, A. S.~G., {et~al.} 2012, MNRAS, 421,
  1007

\bibitem[{{Kennedy} {et~al.}(2015){Kennedy}, {Bamford}, {Baldry},
  {H{\"a}u{\ss}ler}, {Holwerda}, {Hopkins}, {Kelvin}, {Lange}, {Moffett},
  {Popescu}, {Taylor}, {Tuffs}, {Vika}, \& {Vulcani}}]{Kennedy2015}
{Kennedy}, R., {Bamford}, S.~P., {Baldry}, I., {et~al.} 2015, MNRAS, 454, 806

\bibitem[{{Kennedy} {et~al.}(2016){Kennedy}, {Bamford}, {H{\"a}u{\ss}ler},
  {Baldry}, {Bremer}, {Brough}, {Brown}, {Driver}, {Duncan}, {Graham},
  {Holwerda}, {Hopkins}, {Kelvin}, {Lange}, {Phillipps}, {Vika}, \&
  {Vulcani}}]{Kennedy2016}
{Kennedy}, R., {Bamford}, S.~P., {H{\"a}u{\ss}ler}, B., {et~al.} 2016, MNRAS

\bibitem[{{Khochfar} \& {Burkert}(2001)}]{Khochfar2001}
{Khochfar}, S. \& {Burkert}, A. 2001, ApJ, 561, 517

\bibitem[{{Khochfar} \& {Burkert}(2003)}]{Khochfar2003}
{Khochfar}, S. \& {Burkert}, A. 2003, \apjl, 597, L117

\bibitem[{{Kim} \& {Im}(2013)}]{Kim2013}
{Kim}, D. \& {Im}, M. 2013, ApJ, 766, 109

\bibitem[{{Kormendy} \& {Kennicutt}(2004)}]{Kormendy2004}
{Kormendy}, J. \& {Kennicutt}, Jr., R.~C. 2004, ARA\&A, 42, 603

\bibitem[{{La Barbera} \& de~Carvalho(2009)}]{LaBarbera2009}
{La Barbera}, F. \& de~Carvalho, R.~R. 2009, ApJ, 699, L76

\bibitem[{{La Barbera} {et~al.}(2010){La Barbera}, {De Carvalho}, {De La Rosa},
  Gal, Swindle, \& Lopes}]{LaBarbera2010}
{La Barbera}, F., {De Carvalho}, R.~R., {De La Rosa}, I.~G., {et~al.} 2010, AJ,
  140, 1528

\bibitem[{{La Barbera} {et~al.}(2005){La Barbera}, {de Carvalho}, {Gal},
  {Busarello}, {Merluzzi}, {Capaccioli}, \& {Djorgovski}}]{LaBarbera2005}
{La Barbera}, F., {de Carvalho}, R.~R., {Gal}, R.~R., {et~al.} 2005, ApJ, 626,
  L19

\bibitem[{{Lemonias} {et~al.}(2011){Lemonias}, {Schiminovich}, {Thilker},
  {Wyder}, {Martin}, {Seibert}, {Treyer}, {Bianchi}, {Heckman}, {Madore}, \&
  {Rich}}]{Lemonias2011}
{Lemonias}, J.~J., {Schiminovich}, D., {Thilker}, D., {et~al.} 2011, ApJ, 733,
  74

\bibitem[{{Liske} {et~al.}(2015){Liske}, {Baldry}, {Driver}, {Tuffs},
  {Alpaslan}, {Andrae}, {Brough}, {Cluver}, {Grootes}, {Gunawardhana},
  {Kelvin}, {Loveday}, {Robotham}, {Taylor}, {Bamford}, {Bland-Hawthorn},
  {Brown}, {Drinkwater}, {Hopkins}, {Meyer}, {Norberg}, {Peacock}, {Agius},
  {Andrews}, {Bauer}, {Ching}, {Colless}, {Conselice}, {Croom}, {Davies}, {De
  Propris}, {Dunne}, {Eardley}, {Ellis}, {Foster}, {Frenk}, {H{\"a}u{\ss}ler},
  {Holwerda}, {Howlett}, {Ibarra}, {Jarvis}, {Jones}, {Kafle}, {Lacey},
  {Lange}, {Lara-L{\'o}pez}, {L{\'o}pez-S{\'a}nchez}, {Maddox}, {Madore},
  {McNaught-Roberts}, {Moffett}, {Nichol}, {Owers}, {Palamara}, {Penny},
  {Phillipps}, {Pimbblet}, {Popescu}, {Prescott}, {Proctor}, {Sadler},
  {Sansom}, {Seibert}, {Sharp}, {Sutherland}, {V{\'a}zquez-Mata}, {van Kampen},
  {Wilkins}, {Williams}, \& {Wright}}]{Liske2015}
{Liske}, J., {Baldry}, I.~K., {Driver}, S.~P., {et~al.} 2015, MNRAS, 452, 2087

\bibitem[{{Moffett} {et~al.}(2016){Moffett}, {Ingarfield}, {Driver},
  {Robotham}, {Kelvin}, {Lange}, {Me{\v s}tri{\'c}}, {Alpaslan}, {Baldry},
  {Bland-Hawthorn}, {Brough}, {Cluver}, {Davies}, {Holwerda}, {Hopkins},
  {Kafle}, {Kennedy}, {Norberg}, \& {Taylor}}]{Moffett2016}
{Moffett}, A.~J., {Ingarfield}, S.~A., {Driver}, S.~P., {et~al.} 2016, MNRAS,
  457, 1308

\bibitem[{{Oser} {et~al.}(2010){Oser}, {Ostriker}, {Naab}, {Johansson}, \&
  {Burkert}}]{Oser2010}
{Oser}, L., {Ostriker}, J.~P., {Naab}, T., {Johansson}, P.~H., \& {Burkert}, A.
  2010, ApJ, 725, 2312

\bibitem[{Peletier {et~al.}(1990)Peletier, Valentijn, \&
  Jameson}]{Peletier1990}
Peletier, R., Valentijn, E., \& Jameson, R. 1990, A\&A, 233, 62

\bibitem[{{Pipino} {et~al.}(2010){Pipino}, {D'Ercole}, {Chiappini}, \&
  {Matteucci}}]{Pipino2010}
{Pipino}, A., {D'Ercole}, A., {Chiappini}, C., \& {Matteucci}, F. 2010, MNRAS,
  407, 1347

\bibitem[{{S{\'a}nchez-Bl{\'a}zquez} {et~al.}(2014){S{\'a}nchez-Bl{\'a}zquez},
  {Rosales-Ortega}, {M{\'e}ndez-Abreu}, {P{\'e}rez}, {S{\'a}nchez}, {Zibetti},
  {Aguerri}, {Bland-Hawthorn}, {Catal{\'a}n-Torrecilla}, {Cid Fernandes}, {de
  Amorim}, {de Lorenzo-Caceres}, {Falc{\'o}n-Barroso}, {Galazzi},
  {Garc{\'{\i}}a Benito}, {Gil de Paz}, {Gonz{\'a}lez Delgado}, {Husemann},
  {Iglesias-P{\'a}ramo}, {Jungwiert}, {Marino}, {M{\'a}rquez}, {Mast},
  {Mendoza}, {Moll{\'a}}, {Papaderos}, {Ruiz-Lara}, {van de Ven}, {Walcher}, \&
  {Wisotzki}}]{Sanchez2014}
{S{\'a}nchez-Bl{\'a}zquez}, P., {Rosales-Ortega}, F.~F., {M{\'e}ndez-Abreu},
  J., {et~al.} 2014, A\&A, 570, A6

\bibitem[{{Taylor} {et~al.}(2005){Taylor}, {Jansen}, {Windhorst}, {Odewahn}, \&
  {Hibbard}}]{Taylor2005}
{Taylor}, V.~A., {Jansen}, R.~A., {Windhorst}, R.~A., {Odewahn}, S.~C., \&
  {Hibbard}, J.~E. 2005, \apj, 630, 784

\bibitem[{{Tortora} {et~al.}(2010){Tortora}, {Napolitano}, {Cardone},
  {Capaccioli}, {Jetzer}, \& {Molinaro}}]{Tortora2010}
{Tortora}, C., {Napolitano}, N.~R., {Cardone}, V.~F., {et~al.} 2010, MNRAS,
  407, 144

\bibitem[{Vika {et~al.}(2013)Vika, Bamford, H\"{a}ussler, Rojas, Borch, \&
  Nichol}]{Vika2013}
Vika, M., Bamford, S.~P., H\"{a}ussler, B., {et~al.} 2013, MNRAS, 435, 623

\bibitem[{Vulcani {et~al.}(2014)Vulcani, Bamford, Haussler, Vika, Rojas, Agius,
  Baldry, Bauer, Brown, Driver, Graham, Kelvin, Liske, Loveday, Popescu,
  Robotham, \& Tuffs}]{Vulcani2014}
Vulcani, B., Bamford, S.~P., Haussler, B., {et~al.} 2014, MNRAS, 441, 1340

\bibitem[{{Wang} {et~al.}(2011){Wang}, {Kauffmann}, {Overzier}, {Catinella},
  {Schiminovich}, {Heckman}, {Moran}, {Haynes}, {Giovanelli}, \&
  {Kong}}]{Wang2011}
{Wang}, J., {Kauffmann}, G., {Overzier}, R., {et~al.} 2011, \mnras, 412, 1081

\end{thebibliography}

\end{document}